# Spontaneous synchronization and exceptional points in breather complex

Wenchao Wang, [1],[†] Zhifan Fang, [1,2],[†] Tianhao Xian, [1] Mengjie Zhang, [1] Yang Zhao[1], and Li Zhan [1,*]

1, State Key Laboratory of Advanced Optical Communication Systems and Networks, School of Physics and Astronomy, Shanghai Jiao Tong University, Shanghai, China.

2, Zhiyuan honors program, Shanghai Jiao Tong University, Shanghai, China.

† These authors contributed equally to this work.

*Corresponding author. Email: lizhan@sjtu.edu.cn

## Abstract

We experimentally demonstrate the spontaneous synchronization and the exceptional point (EP) induced pulse generation mechanism in the breather complex. The breathing frequency and phase are found to be synchronized during the formation of a 9-breather assembled complex in a mode-locked fiber laser. The breathers are formed at exactly the time point of the complex's breathing frequency leaving or entering the subharmonic entrainment. Such new pulse generation mechanism should be related to the non-Hermitian EPs or the time translation symmetry breaking. The investigations of destroying and rebuilding the mode-locking reveal the connection between the synchronization and laser stabilization. These findings may inspire a wide range of researches including ultrafast optics, micro-cavity combs, ocean breather behaviors, non-Hermitian optics, etc.

## Introduction

Soliton is usually born in dispersive Kerr mediums, mathematically described as a stationary solution of nonlinear Schrödinger equation (NLSE) [1]. It has attracted many interests in different areas like hydrodynamics [2], biology [3] and marine science [4] in past decades. The loss of stability [5-7] is an eternal theme in nonlinear physics, and surely it should be common and widespread in dispersive Kerr mediums. One of oscillation instability of equilibrium called breather soliton receives much attention [8-17], in which energy concentrates in a localized and oscillatory fashion. Mathematically, they are periodic solutions of NLSE, and physically, they can live in the form of ocean behaviors [8-10] or optical pulsations [11-17]. Breathers are attractive in both practical applications [15-17] and fundamental importance in nonlinear science [18]. For instance, they change periodically in both pulse energy and spectral shape, characterized by a breathing frequency $f_b$, and this extra frequency brings new physics. In 2019, Cole theoretically predicted the subharmonic entrainment (SHE) of Kerr breathers [19]. That is the fineness-dependent $f_b$ can be synchronized to the repetition rate $f_r$ in a Kerr resonator, and it forms a frequency-locking plateau of $f_b$ irrelevant to the fineness. What's more, SHE breaks the continuous time-translation symmetry (TTS) [19, 20]. TTS is equivalent to the conservation of energy according to Noether's theorem [21], and the breaking of TTS can create an emerging concept called 'time crystal' [22]. Although there is controversy accepted as real time crystals [23-25], mainstreams commonly admit SHE breather have time-crystalline signatures [26]. Soon later, our group experimentally verified the prediction via a mode-locked fiber laser [27]. Besides, we numerically and theoretically establish the contact to the exceptional points (EPs) and non-Hermitian optics [28] for further demonstration.

On another point, SHE of breathers can be classified as a generalized synchronization, namely, one harmonic of $f_b$ is synchronized by the cavity frequency $f_r$. Synchronization is that parts interact with each other and promote the whole system to form a unified rhythm and phase, and it exists everywhere in real world [29-31]. Under intracavity circumstances of strong nonlinearity or high gain, multiple solitons should be generated [32]. Breathing phenomenon is related to the dynamic loss of stability. Naturally there should existing breathing, or a collective instability of multiple pulses under strong nonlinearity backgrounds, in follow-up we define them as breather complex. This brings bunches of breathing frequencies, and it also inspire new physics. For instance, we would ask, can the synchronization or other mutual interactions among the multiple breather frequencies exist? will such interactions involve with the laser stabilization? How does the SHE influence the breather complexes? It is challenging but crucial to figure out how these extra frequencies change the complex's behaviors.

In this work, we track the formation of the breather complex and capture the spontaneous synchronization in the breather complex, including both the frequency and phase aspects. We use the dispersive Fourier transformation (DFT) [33] to monitor the evolution, and integrate the single-shot spectra to get the pulse energy variation. During the complex formation, each newly formed breather is born exactly at the EPs of the complex's SHE state. At last, we change the intra-cavity transmittance to destroy and rebuild mode-locking. The noise-like signals evolve towards the regularity and eventually become stable accompanied by the synchronization of $f_b$. We believe similar mechanisms may exist widely in optical cavities, ocean dynamics, and otherwise.

## **Results**

**The concept of spontaneous synchronization in breather complex.**

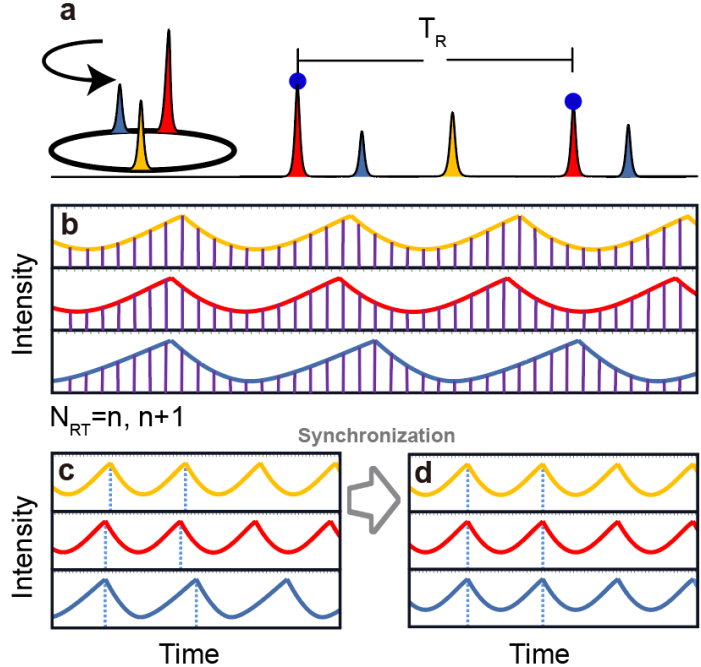

**Fig. 1. Conceptual model.** (**a**) A complex state composes evolution of three pulses at each roundtrip. (**b**) Each independent evolved pulses may breathe at different rhythms, such as the inconsistent phase (see the red and yellow cases), or mismatched frequency (see the red and blue cases). (**c-d**) The mismatches can be eliminated by the spontaneous synchronization effect.

Here, we discuss the breather complex, which consisting of independent breathers with disparate and varied pulse energies within each roundtrip. For the demonstration, we draw a three-pulse structure (each marked by different colors) in Fig. 1a. These breathers have their own breathing modes (see Fig. 1b, the three breathers have different breathing period and phase). Obviously, these breathing patterns should keep and operate independently if there is no mutual interaction. But our argument is that, they can be synchronized to each other by the spontaneous synchronization effect, embodied in both aspects of phase (see yellow and red cases in Figs. 1(c, d)) and frequency synchronization (see red and blue cases in Figs. 1(c, d), and finally the synchronization promotes the initial breathing modes to generate a complex with a uniform breathing rhythm.

**Experimental demonstration of the phase and frequency synchronization.**

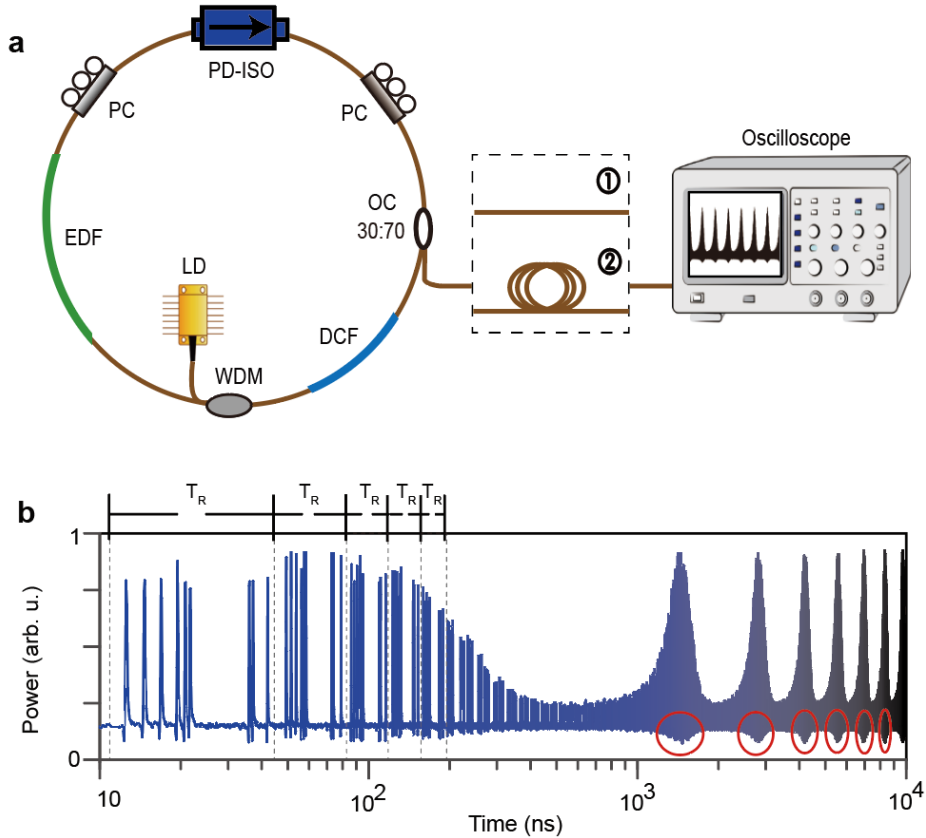

**Fig. 2. Experimental setup and the breather complex formation.** (**a**) The left part is the mode-locked fiber laser and the right part is the direct recording and the time-stretch scheme for real-time detection. (**b**) Oscilloscope trace of the breather complex.

A typical dispersion managed fiber laser [34] is employed to explore the proposed demonstration, sketched in Fig. 2a. The laser is mode-locked by the nonlinear polarization rotation (NPR) technique [35], incorporates a 2 m Er-doped fiber (EDF), and pumped by a 980 nm laser. The section of 40-cm-long dispersion compensating fiber (DCF) is equipped for cavity dispersion management. The other fibers are the standard single mode fibers (SMF), and the total cavity length is 7.65 m, corresponding to a repetition rate of 26.92 MHz. The group velocity dispersion is 65 ps$^2$ km$^{-1}$ for EDF, -48 ps$^2$ km$^{-1}$ for SMF, and −22 ps$^2$ km$^{-1}$ for SMF at 1550 nm. The NPR mode-locking is achieved by a polarization dependent isolator (PD-ISO) and two polarization controllers (PCs). For further characterizing the complex evolution, we take two schemes in the follow-up measurements. One is directly recording the transient process by triggering the oscilloscope, and it can provide evolutionary information from time domain; The other is the time-stretch scheme by using an extra 2km-long SMF, as shown in Fig. 2a. This cannot give the single-shot spectra, because in multi-pulse cases, the adequate dispersion for identifying spectral shape will make the temporal waveforms overlap. We use the scheme mainly for calculating the pulse energy. By integrating the normalized stretched temporal waveforms, we can check how the pulse energy varies during the evolutions.

We find a stable complex mode-locking state at the pump power of 600mW, as shown in Fig. 2b. The time coordinates are logarithmically constructed for displaying the oscilloscope trace of the complex. It has a constant breathing period of ~2μs, and each roundtrip (RT) contains 9 pulses. The waveform shape of the complex resembles the Q-switched mode-locking [36], but the breathing period (which is one order of magnitude smaller than the lifetime of the occupied level in $Er^{3+}$) eliminates the possibility. This 9-pulse structure support self-start, and can maintain the mode-locking for several hours. The judgment that the complex is composed of breathers comes from not only the oscillating appearance of the pulse waveform, but also that, all of the minimum intensity emerge at each peak positions (see the red marks in Fig. 2b). In breathers or peregrine solitons [18], there are featured intensity dips (lower than the background intensity) distributed on both sides of the pulses at each breathing peak.

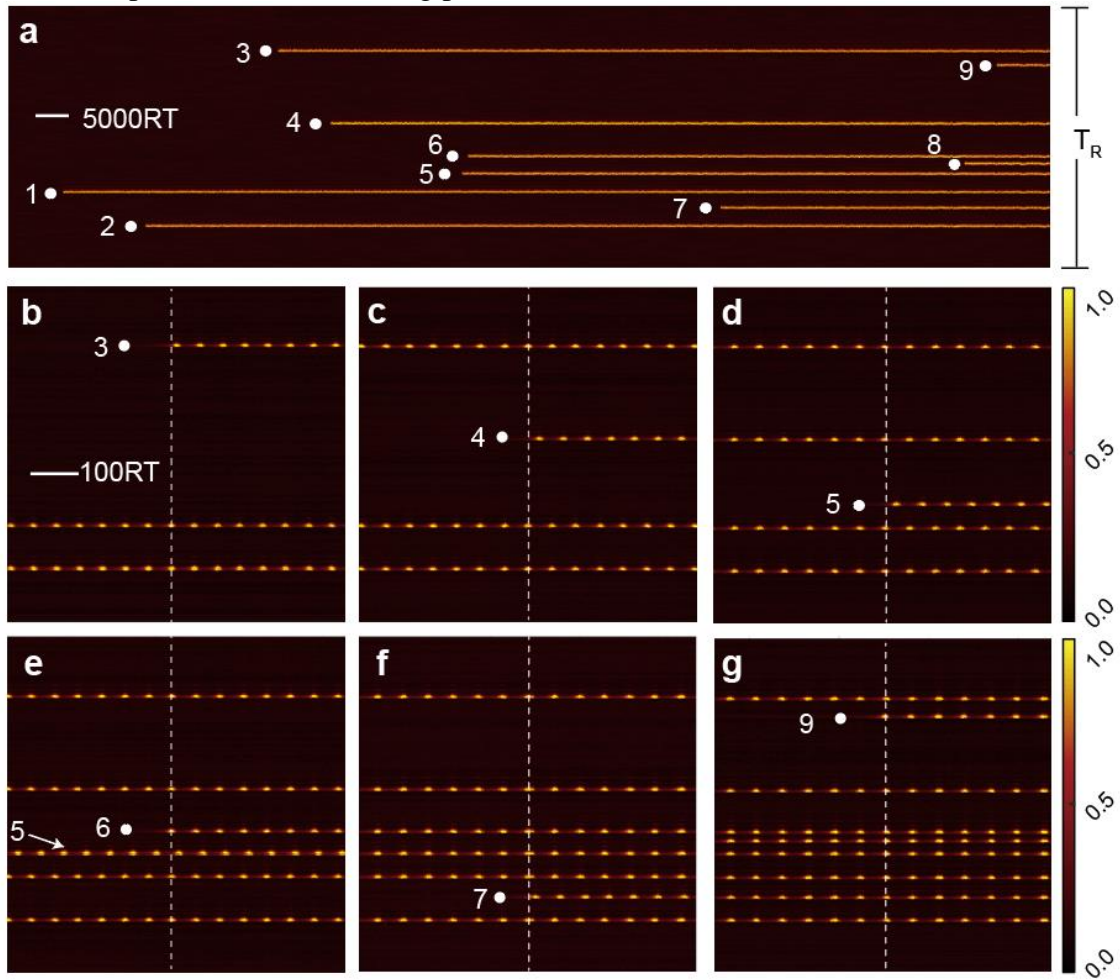

**Fig. 3. Phase synchronization.** (**a**) Shot-to-shot build-up process of the 9-breather complex, recording 160,000 RTs in total (the roundtrip time $T_R$=37.1ns, corresponding to the repetition rate of 26.92 MHz). (**b-g**) Selected local details (each contains 800 RTs), including the cases marked by 3-7, 9 in (**a**). The color bar scales the intensity of the temporal waveforms.

Breathing period depends on the cavity fineness, which is sensitive to the gain/loss [19]. The newly born pulses should suffer from instable gain at first, because soliton or pulse trains are inspired by the modulation instability [1]. Consequently, each new breather should have varied and instable breathing period during the initial time. However, the captured 9 breathers in Fig. 2b have a unified breathing rhythm, indicating the existing of the synchronization. For verification, we track the birth of the complex by the first scheme in Fig. 2a. The recorded time series is segmented with respect to the roundtrip time and displays the formation process in Fig. 3. Nine breathers emerge one by one, they have similar pulse intensities, but the occurrence of sequence in time domain is random, as shown in Fig. 3a.

We list 6 selected local details of the formation process in Figs. 3(b-g) to reveal how the synchronization changes the breathers' evolution. From the diagrams, the breathing periods at initial stages are close, but the phase of breathing envelope are varied. For instance, the initial phase of each newly born breather in Figs. 3(b-f) lags behind the rhythm of the whole, while in Fig. 3g the phase is slightly in advance. After hundreds of roundtrips, the breather complex eventually forms a unified breathing no matter what these initially formed paces are, verifying the phase synchronization. It seems that no extra nonlinear interactions exist among the complex, because the movement of each breather are independent, no collisions and annihilations. We believe the spontaneous synchronization comes from the coupling of breathing modes. These modes are close to each other in both frequency and intensity aspects, and they receive feedbacks from the same cavity. Then the synchronization can spontaneously happen just like that the two asynchronous clocks hanging on the wall will eventually synchronized. It is a universal in the real world, and can be well explained by the Kuramoto model [37] (see methods). The breathing behaviors are coupled by each other, then eliminate the initial phase deviation, and finally realize the synchronization.

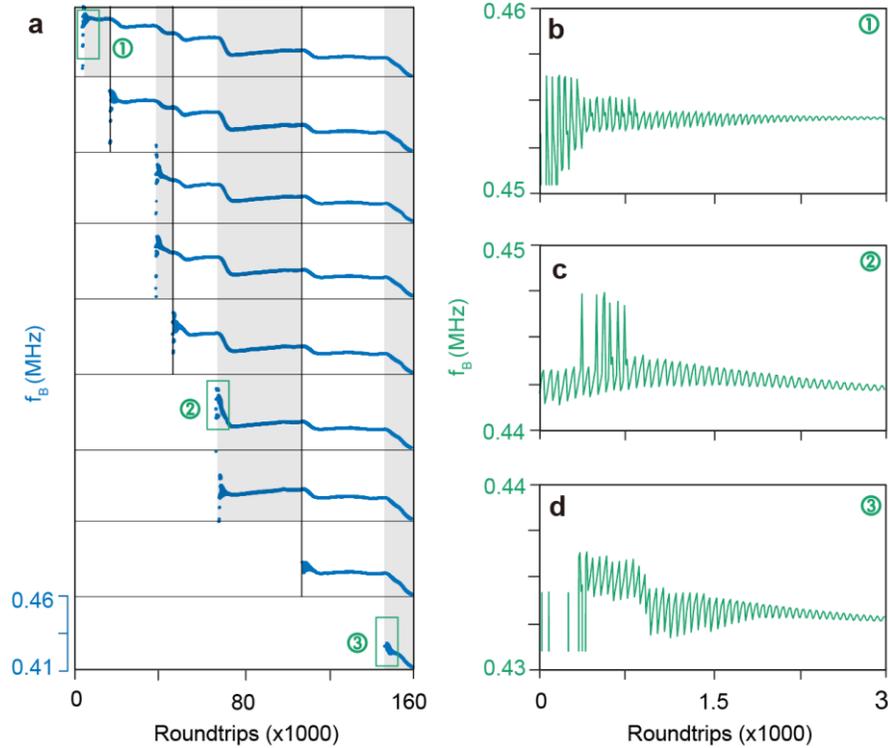

**Fig. 4. Frequency synchronization.** (**a**) The breathing frequency variation of each breathing mode during the build-up process. (**b-d**) Local details of the initial oscillatory behavior, corresponding to the green boxes in (**a**).

We also calculate $f_b$ of each breathing mode (see methods) during the start-up evolution, as shown in Fig. 4a. All of the 9 breathers have the same variation on breathing frequency. It corroborates the claim of the frequency synchronization, and we can also explain this regularity by the coupling theory (see methods). Besides, each emergence of the new breather always leads to a decrease in $f_b$ of other breathers. Under the existing presupposition, $f_b$ depends on the cavity gain, and the gain is sensitive to the pulse energy. The synchronization can promote a unified $f_b$, so the effect may regulate the pulse energy of the 9 breathers to be equal. The result also shows an intriguing and counter-intuitive phenomenon: $f_b$ of all the new generated breathers show the same decayed oscillations at the initial stage. The scatter diagram in Fig. 4a cannot intuitively display the phenomenon, so we depict the line diagrams of three selected local details in Figs. 4(b-d), corresponding the number marks in Fig. 4a. These frequency oscillations should not be aroused by the synchronization, because the pattern has appeared during the birth of the first breather, and in this case there are no extra breathing modes existing to support a synchronization.

**Subharmonic entrainment and new breather generation.**

It seems that $f_b$ is pulled by a kind of force or potential according to the decayed frequency oscillations, and such similar behaviors indicates the same physical origin. One possibility is the SHE effect, which is a synchronization between $f_b$ and cavity frequency $f_r$, and has been proved exert great impact in single breather cases. Since the frequency data acquisition in Fig. 4 comes from the Fourier transformation, each value is calculated by a large number samples, so it causes the results averaged. For the verification, we track the build-up again by the second scheme in Fig. 2, attempting to reveal the instantaneous frequency by analyzing the pulse energy variations. Figure 5a shows the scatter diagram of the total pulse energy changes within each roundtrip. As an illustration, each new breather generation can be distinguished by the 9 bulges. The line graphs in Fig. 5b comes from the same original data in Fig. 5a, the difference is that we draw the diagrams by step $m$ ($m$ =55~59) RTs. This is a kind of down conversion display. Once the complex reaching SHE state, the energy curve should close to straight line, while the non-SHE breathers cannot conform to this rule (see methods). From comparison of Figs. 5 (a, b), we find all of the breathers generated right at the timing of the complex entering or leaving the SHE state (marked by serial numbers of 1~9). This confirms the conjecture, that is, the formation of the breather complex is highly correlated with the SHE effect. Although no new breathers were produced when leaving the last SHE state (marked by number 10 in the last graph of Fig. 5b), the energy of the complex start to decrease, acknowledging the importance of SHE to the complex's evolution and stabilization.

In framework of NLSE modeling, the solitons are induced by the modulation instability [1], from a certain shape wave at a random time. That means the occurrence of the new pulses should certainly be stochastic. However, the breathers in Fig. 4 emerge at just the EPs, or more specifically the critical points of breaking or establishing the complex's SHE state. So, these EPs must change some certain system factors significantly. To respond to the call of 'inverse Occam's razor' [38], here we offer possible explanations. Recent reports show that EPs in non-Hermitian optics are quite sensitive to the disturbance, and hence can be used to enhance sensing [39-41]. It has been proved that the SHE effect arises between the two EPs, and the SHE breathers are right the non-Hermitian degenerates [27]. Hence the energy fluctuations near EPs should be amplified. It creates a high instable background, and naturally new pulses are more likely to occur. On the other hand, SHE breathers present the spontaneous violation of the discrete time-translation symmetry (for $t \rightarrow t + nT_R, n = 1, 2, ...$). So the SHE state indeed breaks the TTS, and the mainstreams also admit the SHE breathers as having the time-crystalline signatures. Here the new SHE breathers are constantly produced during TTS breaking. It's known that spatial translational symmetry (STS) underlies the formation of crystals and the phase transition from liquid to solid [23]. We wonder that the breather's formation could be an analogy to the STS promoted crystallization in spatial crystals [42].

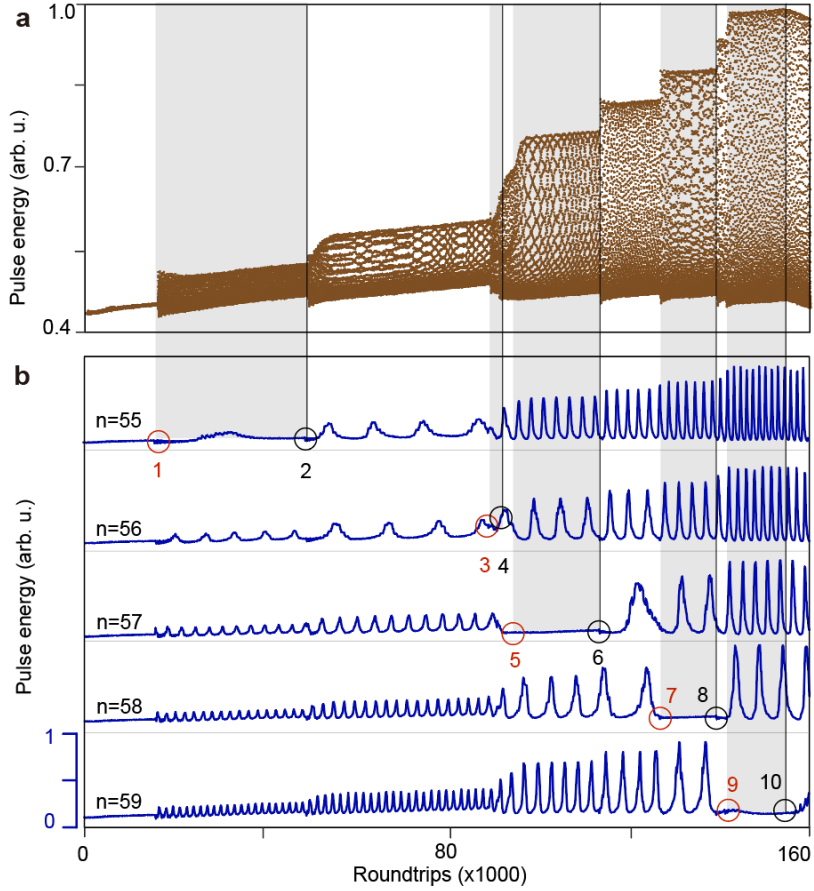

**Fig. 5. Experimental verification of the EP induced pulse generation mechanism.** (**a**) The pulse energy variation during the formation process of the complex, calculated by integrating the single-shot spectra. (**b**) Pulse energy variation displayed by the down conversion method.

**Synchronization, SHE effect and laser stabilization.**

Moreover, we record the oscilloscope traces at different PC angles to study how the effects influence the laser stabilization. The angles are turned towards the same direction from Fig. 6a to 6f, and Figs. 6(g-l) show the corresponding waveforms. It is equivalent to change the intra-cavity transmittance curve, and the mode-locking are destroyed and rebuilt during the process. The synchronization may not be spontaneous achieved at all times. For instance, breathers are unsynchronized in Figs. 6 (a, b), and consequently, their oscilloscope traces in Figs. 6(g, h) are disordered. The variation of the PC angles changes the transmittance curve, also the cavity fineness, so the $f_b$ in these cases are different. According to the discussion in Fig. 5, we suspect that the new breather generation should also be related to the EPs or TTS breaking. The complex is synchronized in Fig. 6c, and its waveform shows a stable breathing (Fig. 6i). Further rotation of PCs decays the breathing depth in Fig. 6(d-f, or j-l). The disappear of breathing pattern is common in single breather cases (usually by turning the pump power). Here the complex showing no breathing properties and finally became stable multiple pulses in Fig. 6f. Its known that the multiple pulses can change to the stationary single pulse state, so the proposed mechanisms should be essential for laser stabilization.

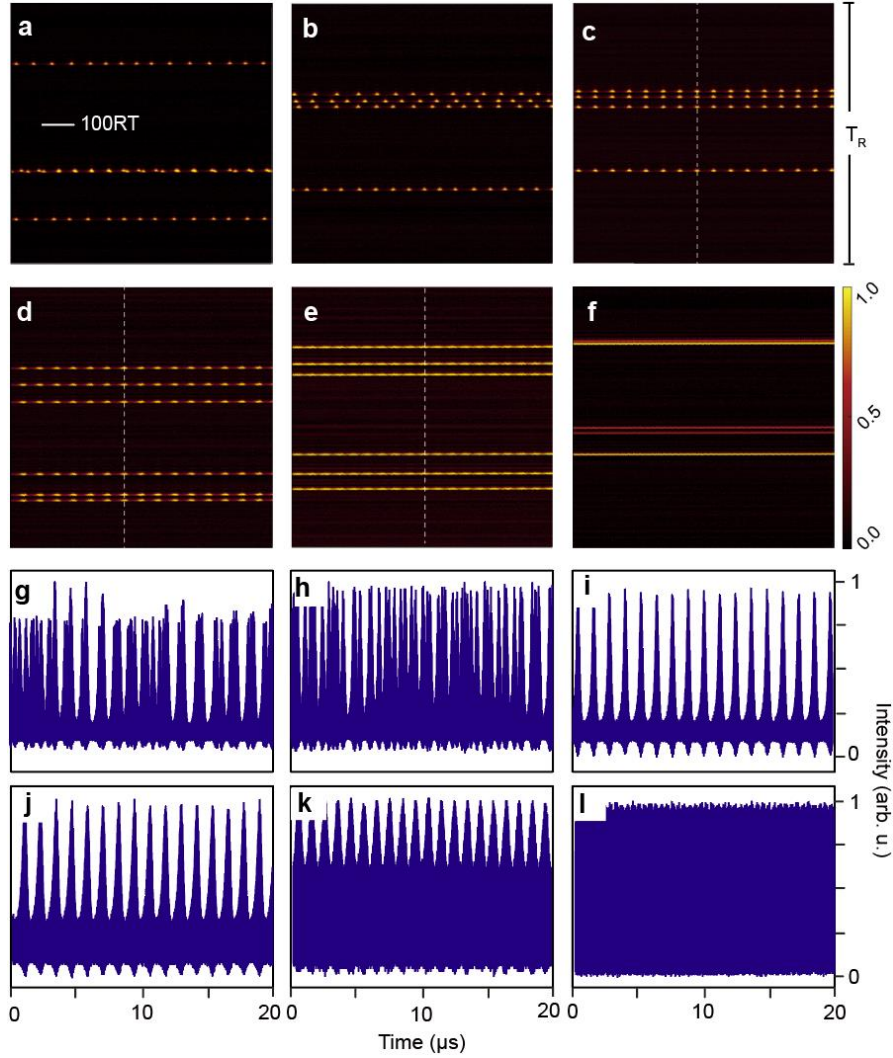

**Fig. 6. Experimental results at different PC angles.** (**a-f**) Shot to shot display. (**g-l**) The corresponding oscilloscope traces.

In conclusion, we report the synchronization and the EP effect in a complex that compose 9 breathers by real-time tracking the build-up process. Both the breathing phase and frequency of each newly formed breathers are spontaneously synchronized by the complex within subsequent hundreds of RTs. Moreover, all of the new breathers are generated exactly when the complex entering or leaving the SHE state. The initial decayed frequency oscillations of $f_b$ proves the strong connection between the pulse generation and the SHE effect. We believe this exotic phenomenon should be related to the exceptional points effect in non-Hermitian optics, or the TTS breaking. The investigations at different PC angles reveal that the unsynchronized multiple breathers exhibit chaotic waveforms, while the synchronized complex can be stabilized to multi pulses. These revealed mechanisms may be benefit many areas including ultrafast laser, breather dynamics, micro resonator, and non-Hermitian optics.

## Methods
### Phase synchronization by the Kuramoto model

Kuramoto model can explain the spontaneous phase synchronization of nonlinear oscillators in many cases. Here, the breather complex shows similar behavior as the coupling among nonlinear oscillators. The breathing envelope are coupled to all of others, just like the intrinsic natural frequency in each single oscillator coupling to all other oscillators. So the breathing phase follows the below governing equations:

$$\frac{d\theta_i}{dt} = f_b + \frac{K}{N}\sum_{j=1}^{N}\sin(\theta_j - \theta_i), \qquad i = 1\ldots N, \qquad (1)$$

where the complex is composed of $N$ breathers (in our case, $N=9$), $\theta_i$ and $\theta_j$ represent the phase of two independent breather, $K$ is the coupling constant. Eq. (1) can only converge at $\theta_j = \theta_i$, that is, all of the breathers will eventually achieve the phase synchronization during the evolution.

### Breathing frequency calculation of the breathing mode in Fig. 3

For characterizing the frequency of each independent breathing mode, we need to separate the breathing modes from the raw oscilloscope waveform data firstly. The repetition rate $f_r$ of the signals can be acquired by the discrete Fourier transform. Then the original data are cut to millions of successive RTs according to $f_r$ and it gives the shot-to-shot evolution shown in Fig. 4. The positions of different breathing modes are fixed within each roundtrip, and each mode occupied 8 sampling points (the sampling rate is 10G sample/s, and corresponds to 370 data points in each RT). We calculate each $f_b$ of breathing mode follow the same steps: Finding out the maximum of these 8 effective data, then we get the approximate breathing envelope of each mode, and finally, with the Fourier transform we can get the exact value of $f_b$.

**Frequency synchronization by the coupling theory**
For simplicity, we demonstrate the effect by a two-frequency model. The coupled equation can be written as:

$$\frac{d}{dt}\begin{pmatrix}a_1\\a_2\end{pmatrix}=-2\pi i\begin{pmatrix}f_{b1} & i\eta\\ i\eta & f_{b2}\end{pmatrix}\begin{pmatrix}a_1\\a_2\end{pmatrix} \quad (4)$$

where, $\eta$ is the coupling efficient, $a_{1,2}$ are the amplitudes of the two breathers, and $f_{b1}, f_{b2}$ are the breathing frequency of the two breathers, $t$ represents the evolution time. The equation has eigenvalues of $f_{1,2}=f_{ave}\pm\sqrt{f_{diff}^2-\eta^2}$, and $f_{ave}=(f_{b1}+f_{b2})/2$, $f_{diff}=(f_{10}-f_{20})/2$. The eigenvalues are the respective frequency of the two breather. Thus we can get the relation $f_1=f_2+2\text{sgn}(f_{diff})\sqrt{f_{diff}^2-\eta^2}$. When $f_{diff}\leq\pm\eta$, the two eigenvalues coalesce and $f_1=f_2$. If $f_1$ and $f_2$ are close and the coupling effect is stronger than the separate effect caused by the frequency difference, the frequency synchronization can be realized.

**Down conversion display in Fig. 5**
Assuming $f_b=mf_r$, the pulse energy in $n$ th RT should be equal to (or very close to) that in $n+m$, $n+2m$ ..., $n+km$ RT ($k=3,4,5....$), and then the energy curve which displayed by stepping each $m$ RT should close to straight line. Because taking the abscissa step $m$ RT is equivalent to draw the curve $y'=\sin[(f_b-mf_r)T]$ instead of $y=\sin(f_bT)$. So it shows a straight line in complex's SHE state in Fig. 5. But in cases of non-SHE breathers, $f_b\neq mf_r$, and the energy curve is $y'=\sin(\Delta fT)$. The farther away the complex get from SHE state, the greater the oscillation frequency $\Delta f$ will be in this display mode.

**Acknowledgments:**

National Natural Science Foundation of China grant 11874040 (PV, CHO). Foundation for leading talents of Minhang, Shanghai. The China Postdoctoral Science Foundation under the Grant 2021M702149.

**Author contributions:**

L. Z., and W.W. designed and interpreted the results of all experiments. W. W. and Z. F. performed the experiments, analyzed the results. T. X., M. Z., and Y. Z. provided advices. W.W. wrote the paper. L. Z. review and editing the paper.

**Competing interests:**

Authors declare that they have no competing interests.

**Materials & Correspondence:**

Correspondence and requests for materials should be addressed to L. Z.

**Data availability:**

All data in the main text or the supplementary materials are available from the corresponding authors on reasonable request. Source data can be provided with this paper.